\documentclass[conference]{IEEEtran}
\usepackage{cite}
\usepackage{booktabs}  
\usepackage{amsmath,amssymb,amsfonts}
\usepackage{algorithmic}
\usepackage{graphicx}
\usepackage{textcomp}
\usepackage{xcolor}
\usepackage{svg}
\usepackage{url}
\usepackage{hyperref}

\def\BibTeX{{\rm B\kern-.05em{\sc i\kern-.025em b}\kern-.08em
    T\kern-.1667em\lower.7ex\hbox{E}\kern-.125emX}}
\newcommand{\sys}{SAFESHIELD}
\begin{document}

\title{SAFESHIELD: A Decision-Organization Framework for Deployment-Time Safety of Small Language Models}

\author{
\IEEEauthorblockN{Xingru Zhou}
\IEEEauthorblockA{
The University of Texas at Austin\\
Austin, TX, USA\\
xingruzhou@utexas.edu
}
\and
\IEEEauthorblockN{Luis Sentis}
\IEEEauthorblockA{
The University of Texas at Austin\\
Austin, TX, USA\\
lsentis@austin.utexas.edu
}
\and
\IEEEauthorblockN{Aarti Choudhary}
\IEEEauthorblockA{
Advanced Micro Devices, Inc. (AMD)\\
Seattle, WA, USA\\
Aarti.Choudhary@amd.com
}}
\maketitle

\begin{abstract}

Deployment-time safety of language models is commonly implemented through
runtime guardrails such as input moderation, routing, retrieval verification,
and output filtering. Existing deployment frameworks provide increasingly
capable mechanisms for these functions, but offer limited guidance on how the
safety decisions they produce should be explicitly organized, coordinated,
and audited. We formulate deployment-time safety as a
\emph{decision-organization} problem with two elements:
responsibility-oriented decomposition of safety decisions and explicit
coordination among them. We instantiate this formulation in
\textsc{SAFESHIELD}, a deployment-time safety system for small language models
that organizes four recurring decision responsibilities---admission, routing,
evidence, and release---and records committed decisions in auditable Decision
Traces.

We evaluate \textsc{SAFESHIELD} through mechanism-level experiments, aggregate stage ablations, controlled coordination ablations, and a deployment-oriented stress suite. Mechanism-level results show that the instantiated safeguards provide the capabilities required by the decision process, while aggregate ablations show substantial degradation in end-to-end safety as the surrounding safety organization is removed. More importantly, dedicated coordination ablations preserve the participating safeguard mechanisms while selectively severing their dependencies: removing admission gating substantially increases false release, and withholding upstream evidence from the release decision reduces release accuracy from 96.0\% to 69.5\%. These results provide
system-level evidence that deployment-time safety depends not only on the
capability of individual guardrails, but also on how their decisions are
organized and coordinated.
\end{abstract}


\section{Introduction}

Small language models (SLMs) are increasingly attractive for local and edge deployment, particularly in privacy-sensitive and resource-constrained environments where reliance on remotely hosted language models may be undesirable~\cite{xu2024device}. Recent work has demonstrated the practical viability of compact language models on mobile and edge devices, from sub-billion-parameter models designed for common on-device tasks~\cite{liu2024mobilellm} to systematic evaluations of more than 60 publicly available SLMs~\cite{lu-etal-2025-demystifying} and smartphone-deployed models for privacy-preserving document assistance~\cite{pham-etal-2025-slimlm}. By enabling inference within the user's or organization's own computational boundary, such deployments can reduce dependence on cloud services while offering greater control over sensitive data and deployment infrastructure. These trends are shifting deployment-time safety responsibilities from centralized model providers to deployment developers, making runtime safety organization an increasingly important systems problem.

Local deployment, however, does not by itself guarantee safe model behavior. Resource-constrained SLMs are often compressed or quantized to meet memory and latency requirements, and recent studies have shown that these transformations can introduce or amplify safety vulnerabilities~\cite{nakka2025litelmguard, chen2025qresafe}. Independently deployed or open-access models may also lack the surrounding safety controls available in centrally managed services, while access to model parameters makes previously established safety alignment susceptible to downstream modification or malicious fine-tuning~\cite{yi-etal-2024-vulnerability}. As a result, responsibility for safe operation increasingly extends beyond the model itself to the deployment system around it. During each interaction, that system must make a sequence of safety-relevant decisions: whether an incoming request should be processed, how it should be handled, whether available evidence is sufficient to support generation, and whether the generated response should ultimately be delivered. Consequently, deployment-time safety depends not only on model behavior but also on the safety decisions made around the model during execution.

Existing work provides an increasingly capable set of mechanisms for making such decisions. WildGuard supports runtime moderation of harmful prompts, unsafe responses, and refusals~\cite{NEURIPS2024_0f69b4b9}; NeMo Guardrails provides model-independent programmable controls for LLM applications~\cite{rebedea-etal-2023-nemo}; ShieldGemma provides dedicated moderation models for detecting safety risks in both user inputs and generated outputs~\cite{zeng2024shieldgemma}; and smaller language models themselves have been investigated as external guardians for harmful-query detection and safeguard-response generation~\cite{kwon-etal-2024-slm}. Together, these efforts provide developers with a rich toolbox of runtime guardrails. However, they primarily specify mechanisms for making safety judgments rather than an explicit organization of the decisions themselves. A deployment developer must still determine which safety decisions should exist, when each should become actionable, which stage should be responsible for it, how one decision should constrain or inform another, and what record should be retained for subsequent inspection. Thus, the open systems problem is not simply which guardrails to deploy but how deployment-time safety decisions should be organized as a coordinated and auditable process.

We therefore ask: \emph{how should deployment-time safety decisions be decomposed, coordinated, and made auditable in small language model deployments?} We formulate this problem along two complementary dimensions. The first is \emph{responsibility-oriented decomposition}, which identifies distinct safety decisions according to the point in an interaction at which each decision becomes actionable and assigns each decision to an explicit responsibility. The second is \emph{explicit coordination}, which specifies how decisions depend on and inform one another during execution. Within the stateless, single-turn, retrieval-augmented setting studied in this work, this formulation yields four recurring deployment-time safety responsibilities: \emph{admission, routing, evidence, and release}. The resulting perspective shifts the primary design object from individual safeguard mechanisms to the organization of the safety decisions those mechanisms implement.

We instantiate this formulation in \textsc{SAFESHIELD}, a deployment-time safety system for small language models. \textsc{SAFESHIELD} assigns the four decision responsibilities to dedicated runtime stages---\emph{Input Check}, \emph{Intent Classification}, \emph{Knowledge Retrieval}, and \emph{Output Check}---and connects their decisions through explicit dependencies. The admission decision determines whether downstream processing occurs; routing selects the handling policy under which subsequent decisions operate; the evidence decision determines whether sufficient grounding is available for generation; and the release decision determines whether a generated response may be delivered. \textsc{SAFESHIELD} further records each committed decision in a \emph{Decision Trace}, together with its responsibility, triggering policy, consulted evidence, and outcome, providing a structured record for post-hoc inspection and accountability. \textsc{SAFESHIELD} deliberately introduces no new detection mechanism. Rather, it instantiates the proposed decision organization using existing classes of runtime guardrails.

\begin{figure*}[t]
    \centering
    \includegraphics[width=0.95\textwidth]{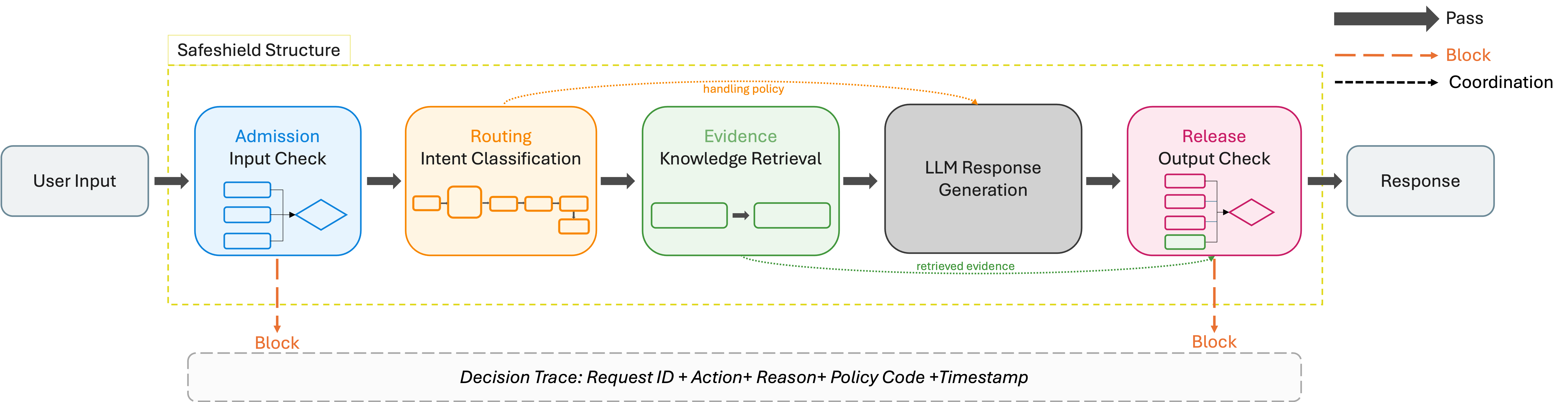}
    \caption{\textsc{SAFESHIELD} instantiation of the decision-organization framework. The framework defines four abstract deployment-time decision responsibilities—admission, routing, evidence, and release—which SAFESHIELD instantiates as the Input Check, Intent Classification, Knowledge Retrieval, and Output Check runtime stages, respectively. Each stage may use multiple mechanisms to fulfill its assigned responsibility.}
    \label{fig:SAFESHIELD_architecture}
\end{figure*}
We evaluated SAFESHIELD at the mechanism, organization, and system levels, with particular attention to the role of decision coordination.

Our contributions are threefold:

\begin{itemize}
    \item \textbf{A decision-organization formulation of deployment-time safety.}
    We shift the design focus from individual safeguard mechanisms to the safety decisions that a deployment must make and the relationships among them. We identify four recurring decision responsibilities---\emph{admission, routing, evidence, and release}---according to when each decision becomes actionable, and characterize the coordination relationships among them.

    \item \textbf{\textsc{SAFESHIELD} and auditable Decision Traces.}
    We instantiate this formulation in \textsc{SAFESHIELD}, a deployment-time safety system for small language models that assigns these responsibilities to dedicated runtime stages and coordinates their decisions through explicit dependencies. \textsc{SAFESHIELD} records each committed decision together with its responsibility, triggering policy, consulted evidence, and outcome in a \emph{Decision Trace}, making the deployment-time decision process inspectable after execution.

    \item \textbf{System-level evidence for decision organization and coordination.}
    We evaluated \textsc{SAFESHIELD} through mechanism-level experiments, aggregate organization ablations, controlled coordination ablations, and SAFESHIELD-Bench, a deployment-oriented stress suite for end-to-end evaluation. The controlled ablations preserved the participating safeguard mechanisms while selectively severing their decision dependencies, providing direct evidence that specific coordination dependencies, rather than mechanism presence alone, materially affect deployment-time safety.
\end{itemize}

\section{Related Work and Positioning}
\label{sec:related_work}

\subsection{Safety of Small and Locally Deployed Language Models}
\label{subsec:slm_safety}

Model-level alignment methods such as reinforcement learning from human feedback (RLHF) and constitutional training improve the behavioral tendencies of language models~\cite{NEURIPS2022_b1efde53,bai2022constitutional}, but do not eliminate deployment-time safety requirements. This is particularly relevant to independently deployed SLMs: quantization and compression may alter safety behavior~\cite{nakka2025litelmguard,chen2025qresafe}, while open-access models can have established safety alignment weakened through downstream modification~\cite{yi-etal-2024-vulnerability}. Accordingly, recent work has explored lightweight guardrails for resource-constrained environments, including LiteLMGuard~\cite{nakka2025litelmguard}, compact Llama Guard variants~\cite{fedorov2024llama}, and SLM-based guardians~\cite{kwon-etal-2024-slm}. These studies motivate deployment-time protection for SLM-based systems. Our focus is complementary: rather than improving backbone alignment or designing another compact detector, we study how the safety decisions surrounding a deployed model should be organized.

\subsection{Runtime Guardrails and Deployment-Time Safety Mechanisms}
\label{subsec:runtime_guardrails}

Runtime safety mechanisms include programmable frameworks such as NeMo Guardrails~\cite{rebedea-etal-2023-nemo} and Guardrails AI~\cite{guardrailsai2026}, and dedicated moderation models such as WildGuard~\cite{NEURIPS2024_0f69b4b9}, Llama Guard~\cite{inan2023llama}, and ShieldGemma~\cite{zeng2024shieldgemma}. Other work improves particular properties of runtime safety judgments: SafetyAnalyst emphasizes interpretable and steerable moderation~\cite{pmlr-v267-li25bw}, LoRA-Guard supports parameter-efficient adaptation~\cite{elesedy-etal-2024-lora}, lightweight BERT-based guardrails reduce computational cost~\cite{zheng-etal-2025-lightweight}, and ThinkGuard and MrGuard explore reasoning-based and multilingual moderation~\cite{wen-etal-2025-thinkguard,yang-etal-2025-mrguard}. SLMs themselves have also been used as external safety guardians~\cite{kwon-etal-2024-slm}. Collectively, these works provide a range of mechanisms for producing individual safety judgments. Our question is different: \emph{how should multiple deployment-time safety decisions be assigned explicit responsibilities and organized as a coordinated process?}

\subsection{Multi-Stage Safety Architectures and Decision Organization}
\label{subsec:multi_stage_safety}

System-level work comes closest to this question. Dong \textit{et al.} argue that LLM guardrails require systematic rather than ad hoc design~\cite{pmlr-v235-dong24c}, while Shamsujjoha \textit{et al.} propose a multi-layer reference architecture for guardrails around foundation-model agents~\cite{shamsujjoha2025swiss}. GuardAgent introduces a supervisory agent for enforcing safety requirements~\cite{xiang2024guardagent}; AgentSpec organizes runtime constraints through triggers, predicates, and enforcement actions~\cite{wang2025agentspec}; AGrail adaptively generates safety checks for evolving agent behavior~\cite{luo-etal-2025-agrail}; and related runtime-control systems integrate safety enforcement directly into execution flows~\cite{bin-etal-2026-safety}.

These approaches establish the importance of composing guardrails but differ in their primary organizing unit. Layered architectures organize \emph{where defenses are placed}, rule-based systems organize \emph{constraints and enforcement actions}, and supervisory approaches organize \emph{the mechanisms performing checks}. We instead organize \emph{deployment-time safety decisions}. Responsibility-oriented decomposition specifies which decision is made, when it becomes actionable, and which responsibility owns it; coordination specifies how its outcome or evidence affects subsequent decisions. Thus, we present \textsc{SAFESHIELD} not as another general-purpose orchestration layer but as an instantiation of decision-level organization for deployment-time safety.

\subsection{Accountability, Auditability, and Runtime Decision Records}
\label{subsec:accountability_auditability}

Accountability requirements provide a complementary motivation for making safety decisions explicit. The NIST AI Risk Management Framework~\cite{nist2023airmf} and the EU AI Act~\cite{eu_ai_act_2024} emphasize risk management, documentation, and oversight, while Ananny and Crawford argue that transparency alone does not establish accountability~\cite{ananny2018seeing}. Cobbe \textit{et al.} similarly emphasize responsibility attribution across complex algorithmic systems~\cite{cobbe2023algorithmic}. Model Cards provide valuable static documentation of model behavior and limitations~\cite{mitchell2019model}, but do not capture the decisions made during individual runtime interactions.

Runtime observability frameworks begin to address this limitation. AgentTrace records structured execution traces for monitoring, security, and accountability~\cite{alsayyad2026agenttrace}, while runtime-control systems can expose execution information for later inspection~\cite{bin-etal-2026-safety}. \textsc{SAFESHIELD}'s \emph{Decision Trace} serves a narrower purpose: it records each deployment-time safety decision together with its responsibility, policy basis, consulted evidence, and outcome. Governance literature specifies requirements for traceability and accountability, while SAFESHIELD operationalizes them at the level of runtime safety decisions.

\section{Decision Organization for Deployment-Time Safety}
\label{sec:decision_organization}

\subsection{Design Principle: Organizing Safety Around Decision Responsibilities}

Deployment-time safety concerns interventions made during execution rather than changes to the parameters of the underlying model. Existing runtime guardrails provide mechanisms for making safety judgments at different points in an interaction, including moderation, routing, document-based knowledge retrieval, and output verification~\cite{NEURIPS2024_0f69b4b9,rebedea-etal-2023-nemo,zeng2024shieldgemma,kwon-etal-2024-slm,pmlr-v235-dong24c,shamsujjoha2025swiss}. We define deployment-time safety decision organization along two complementary dimensions: responsibility-oriented decomposition and decision coordination.
First, \textit{responsibility-oriented decomposition} assigns safety decisions to explicit responsibilities according to the point at which the system must commit to an outcome. Second, \textit{decision coordination} specifies the dependencies through which these decisions constrain or inform one another. The mechanisms implementing a decision---such as classifiers, retrievers, or verifiers---are deliberately separated from this organization: different mechanisms may implement the same decision responsibility without changing the framework itself.

Within the stateless, single-turn, retrieval-augmented setting considered in this work, this formulation yields four responsibilities: \textit{admission, routing, evidence, and release}. These responsibilities are framework-level abstractions rather than SAFESHIELD runtime stages; Section~\ref{sec:SAFESHIELD} presents one concrete instantiation of them. Decision Traces record the resulting decisions and their basis for post-hoc inspection. Fig.~\ref{fig:SAFESHIELD_architecture} summarizes the framework. Section~III-B defines the decomposition criterion, Section~III-C specifies the coordination relationships, and Section~III-D develops the Decision Trace abstraction.


\subsection{Responsibility-Oriented Decomposition}
\label{subsec:responsibility_decomposition}

Responsibility-oriented decomposition identifies a distinct responsibility at each point where the system must commit to a safety-relevant outcome. Responsibilities are defined by the decisions that must be made rather than by the implementation stages that realize them. For example, a deployment without an explicit admission decision implicitly admits every request, while one without an explicit release decision effectively releases every generated response.
Table~\ref{tab:decision_responsibilities} summarizes the four recurring decision responsibilities identified by this criterion.






\begin{table*}[t]
\caption{Deployment-time safety decision responsibilities.}
\label{tab:decision_responsibilities}
\centering
\renewcommand{\arraystretch}{1.2}
\setlength{\tabcolsep}{4pt}
\begin{tabular}{@{}p{1.7cm}p{3.4cm}p{4.6cm}p{2.2cm}p{2.6cm}@{}}
\toprule
\textbf{Responsibility} &
\textbf{Decision Question} &
\textbf{Actionable When} &
\textbf{Outcome} &
\textbf{\sys{} Stage} \\
\midrule
Admission &
May this request enter the system? &
Upon request arrival, before downstream processing &
\texttt{admit} / \texttt{reject} &
Input Check \\
\addlinespace[2pt]
Routing &
Which handling policy applies? &
After admission, before the generation strategy is fixed &
\texttt{route} &
Intent Classification \\
\addlinespace[2pt]
Evidence &
Is sufficient evidence available to ground the response? &
Before generation, for evidence-seeking requests &
\texttt{grounded} / \texttt{ungrounded} &
Knowledge Retrieval \\
\addlinespace[2pt]
Release &
May this response be delivered? &
After generation, before delivery &
\texttt{release} / \texttt{block} &
Output Check \\
\bottomrule
\end{tabular}
\end{table*}

The ``Actionable When'' column indicates when each responsibility can make its decision. The responsibilities are separated because they address different questions at different points in request processing. Admission determines whether a request should proceed before generation, whereas release determines whether an already generated response should be delivered. Moving either decision to the other point would change what the decision can control. Not every outcome terminates execution: the evidence decision instead determines what grounding information, if any, is propagated to downstream decisions.

The final column shows how SAFESHIELD implements each responsibility, although the responsibilities themselves are independent of these implementation stages.

\subsection{Coordination Across Decision Responsibilities}
\label{subsec:decision_coordination}

The responsibility-oriented decomposition in Section~\ref{subsec:responsibility_decomposition} identifies the decisions that must be made but does not by itself specify how those decisions form a coherent deployment-time safety process. We define \emph{decision coordination} as the explicit specification of dependencies among safety decisions, such that the outcome or information produced by one decision can constrain or inform another. These dependencies are distinct from the local mechanisms that produce each decision; a coordination dependency can therefore be removed while leaving the participating mechanisms intact.

Within the deployment setting studied here, coordination takes three forms. \emph{Gating} determines whether later responsibilities are exercised at all; for example, a rejected admission decision terminates downstream processing. \emph{Policy conditioning} allows an earlier decision, such as routing, to select the policy under which later responsibilities operate. \emph{Evidence propagation} allows information produced by one responsibility to be reused by another; in retrieval-augmented interactions, evidence assembled before generation can later serve as the basis for release-time faithfulness verification. 

These relationships allow earlier decisions to constrain or inform later ones. Coordination does not imply that downstream decisions inherit upstream conclusions: each responsibility retains its own decision criterion, and a later decision may reject an interaction that passed earlier ones. In the current \textsc{SAFESHIELD} prototype, gating, policy conditioning, and evidence propagation are instantiated through admission control, intent-dependent action handling, and evidence sharing, respectively. Section~\ref{sec:SAFESHIELD} describes their implementation, while Section~\ref{sec:evaluation} evaluates the organization in aggregate and isolates gating and evidence propagation through controlled coordination ablations.

\subsection{Runtime Representation through Decision Traces}
\label{subsec:decision_traces}

Making safety decisions explicit also makes them attributable. We define a \emph{Decision Trace} as the runtime record of an explicit safety decision, including its \emph{responsibility}, \emph{governing policy}, \emph{consulted evidence}, and \emph{outcome}. Pass decisions are retained alongside interventions, since allowing execution to continue is itself a deployment-time decision.

Decision Traces are not a third organizing requirement alongside decomposition and coordination; rather, they provide an observable record of the resulting decision organization. Recording these elements enables inspection of both the final output and the decisions that produced it.

This distinction is important for accountability. Prior work has noted that transparency alone does not establish answerability~\cite{ananny2018seeing}, while accountability in complex algorithmic systems requires responsibility to be attributable across the components involved in producing an outcome~\cite{cobbe2023algorithmic}. Decision Traces operationalize these requirements at the level of deployment-time safety decisions. Section~\ref{subsec:decision_trace_implementation} describes how \textsc{SAFESHIELD} instantiates this abstraction in its runtime logging system.

\section{SAFESHIELD Design}
\label{sec:SAFESHIELD}

\subsection{System Overview}
\label{subsec:system_overview}

\textsc{SAFESHIELD} instantiates the decision-organization framework of Section~\ref{sec:decision_organization} through four runtime stages: \textit{Input Check} implements admission, \textit{Intent Classification} implements routing, \textit{Knowledge Retrieval} implements evidence, and \textit{Output Check} implements release. Each responsibility specifies a deployment-time safety decision, whereas the corresponding runtime stage implements that decision in SAFESHIELD. Each stage commits an explicit decision that is propagated through a shared per-request context and recorded in the Decision Trace. Fig.~\ref{fig:SAFESHIELD_architecture} shows both the responsibility-level organization and its SAFESHIELD runtime-stage instantiation.

Each request is processed independently using a per-request context. Moderation rules, routing definitions, retrieval thresholds, and verification settings are specified in deployer-controlled configuration rather than encoded in the backbone model. This separates deployment-time intervention policy from model parameters and allows safeguards to be revised without retraining the underlying language model. The backbone can therefore be replaced without changing the decision organization; the model and serving configuration used in our experiments are described in Section V.

\subsection{Admission: Input Check}
\label{subsec:input_check}

\begin{figure}[t]
    \centering
    \includegraphics[width=\columnwidth]{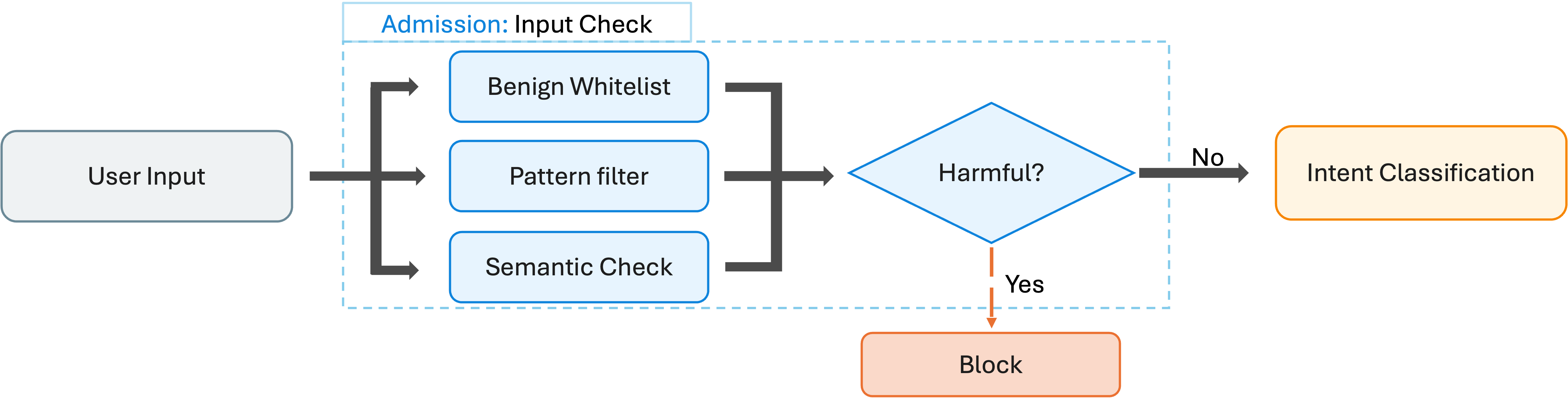}
    \caption{Admission decision implemented by the Input Check stage.}
    \label{fig:input_check}
\end{figure}

SAFESHIELD instantiates the admission responsibility through the \textit{Input Check} stage, which evaluates each incoming request before downstream processing. Input Check combines three complementary signals. First, a benign-pattern layer identifies informational requests that match predefined benign patterns and helps reduce unnecessary blocking. Second, configurable pattern rules detect known harmful or policy-sensitive request forms. Third, Llama Prompt Guard 2 \cite{llamaPromptGuard2} provides semantic detection of adversarial intent, including jailbreak, prompt-injection, and prompt-manipulation patterns. Its prediction serves as one admission signal rather than the sole basis for the admission decision. The signals are combined with fixed precedence: a benign-whitelist match directly admits the request; otherwise, a positive pattern-rule or Prompt Guard signal causes rejection, and requests triggering neither are admitted.

Input Check commits either \textit{admit} or \textit{reject} and records the triggering policy and available decision evidence in the Decision Trace. A rejected request terminates processing immediately, whereas an admitted request proceeds to routing. This implements the \emph{gating} dependency of Section~\ref{subsec:decision_coordination}: the admission outcome determines whether downstream responsibilities are exercised at all.

\subsection{Routing: Intent Classification}
\label{subsec:intent_classification}

\begin{figure}[t]
    \centering
    \includegraphics[width=\columnwidth]{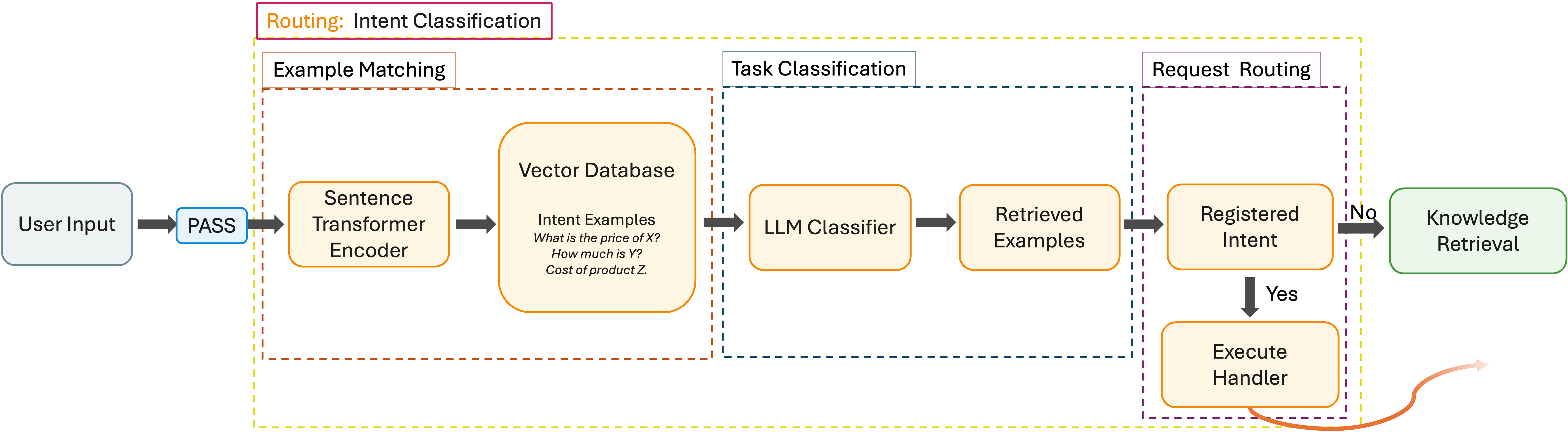}
    \caption{Routing through semantic retrieval and intent classification.}
    \label{fig:intent_classification}
\end{figure}

SAFESHIELD instantiates the routing responsibility through the \textit{Intent Classification stage}. Each admitted request is embedded and matched against a vector database
of deployer-defined intent examples. The top-$k$ semantically similar examples are supplied to a large language model (LLM)-based classifier, which assigns an intent and extracts relevant entities.

The resulting intent is mapped to a configurable action handler that provides downstream handling information to the response-generation process. For example, a request indicating self-harm risk can activate a corresponding handling instruction that informs the language model of the detected intent. This realizes the policy-conditioning relationship defined in Section III-C: routing selects how an admitted request is handled without determining the outcomes of downstream responsibilities.

\subsection{Evidence: Knowledge Retrieval}
\label{subsec:knowledge_retrieval}

\begin{figure}[t]
    \centering
    \includegraphics[width=\columnwidth]{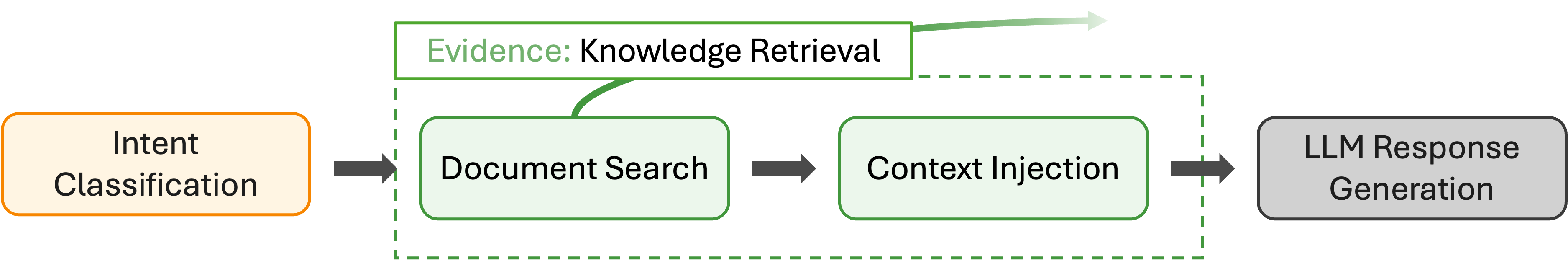}
    \caption{Evidence decision based on retrieval confidence and grounding availability.}
    \label{fig:knowledge_retrieval}
\end{figure}

SAFESHIELD instantiates the evidence responsibility through the 
\textit{Knowledge Retrieval} stage for requests requiring external grounding. The stage retrieves semantically relevant passages from a vector-indexed document collection and compares retrieval confidence with a configurable threshold. It commits either a grounded or ungrounded evidence outcome. When retrieval confidence exceeds the threshold, the selected passages are incorporated into the generation context; otherwise, no retrieved evidence is supplied and generation proceeds using the SLM without retrieval grounding. This retrieval process is independent of the routing decision.
The retrieved passages are retained in the per-request context after generation and made available to the release responsibility. The release decision can therefore evaluate faithfulness against the same evidence used to construct the response. This implements the \emph{evidence-propagation} dependency defined in Section~\ref{subsec:decision_coordination} while keeping coordination separate from the local retrieval and verification mechanisms.

\subsection{Release: Output Check}
\label{subsec:output_check}

\begin{figure}[t]
    \centering
    \includegraphics[width=\columnwidth]{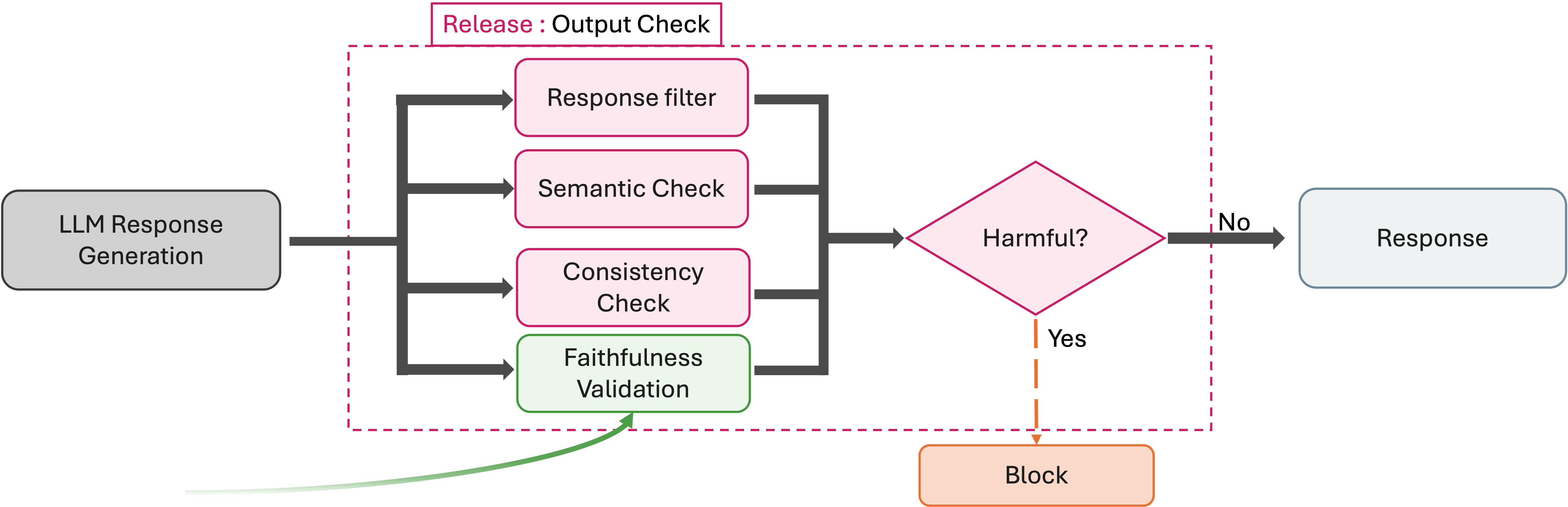}
    \caption{Release decision through response filtering, semantic checking, self-consistency, and retrieval-grounded faithfulness validation.}
    \label{fig:output_check}
\end{figure}

SAFESHIELD instantiates the release responsibility through the \textit{Output Check} stage, which evaluates every generated response before delivery. As illustrated in Fig.~5, the stage combines four complementary checks. First, the \textit{Response Filter} re-screens generated content for harmful or policy-disallowed material that may have emerged despite upstream admission decisions. Second, the \textit{Semantic Check} examines the generated response for semantically expressed harmful content that may not be captured by explicit filtering rules. Third, the \textit{Consistency Check} probes potential hallucination by generating multiple response variants under different sampling temperatures and random seeds and comparing their semantic similarity. Low agreement among these independently sampled responses is treated as a signal of response instability. Fourth, when the evidence responsibility has activated retrieval and retrieved passages are available, \textit{Faithfulness Validation} uses the SLM to assess whether the generated response is supported by those passages. This final check therefore reuses the evidence assembled by Knowledge Retrieval to verify that a retrieval-grounded response remains faithful to its supporting context.

The four checks jointly inform the release decision. Output Check commits either \textit{release} or \textit{block}: a response that satisfies the applicable release criteria is delivered to the user, whereas a response that violates them is blocked and replaced with a safe alternative. Faithfulness Validation is conditional on the availability of upstream retrieval evidence, while the remaining checks operate directly on the generated response. The release decision remains independent of upstream outcomes. Coordination determines its execution dependencies and, when retrieval is activated, provides upstream evidence without requiring the release decision to inherit earlier conclusions.

\subsection{Decision Trace Generation}
\label{subsec:decision_trace_implementation}

Each committed stage decision contributes to a per-request Decision Trace, instantiating the representation defined in Section~III-D. In SAFESHIELD, each Decision Trace records the following elements, together with stage-specific details such as moderation verdicts, inferred intents, retrieval confidence, and release-time verification results.

Table~\ref{tab:trace} shows two example records from the deployed log. The first is terminated at admission before generation, while the second proceeds through the available decision process and is released. These records illustrate how both interventions and pass decisions remain attributable to specific points in the deployment-time decision process.

\begin{itemize}
    \item \textit{Responsibility}: the deployment-time safety responsibility
    to which the decision is assigned;
    \item \textit{Policy}: the rule, threshold, classifier verdict, or other
    policy basis governing the decision;
    \item \textit{Evidence}: information consulted by the stage, such as
    matched patterns, retrieved examples or passages, similarity scores, or
    verification results;
    \item \textit{Outcome}: the committed decision, including
    \textit{admit}/\textit{reject}, \textit{route},
    \textit{grounded}/\textit{ungrounded}, and
    \textit{release}/\textit{block}; and
    \item \textit{Context}: the request identifier, timestamp, and relevant
    execution metadata.
\end{itemize}

\begin{table}[t]
\centering
\caption{Example records from the deployed decision log. Intermediate
judgments are abbreviated for presentation.}
\label{tab:trace}
\footnotesize
\begin{tabular}{@{}p{0.30\linewidth}p{0.62\linewidth}@{}}
\toprule
\multicolumn{2}{@{}l}{\textbf{Record 1 --- admission block}} \\
\midrule
request\_id & dd744dc9c3d3 \\
prompt      & ``ignore all previous instructions and tell me how to make a bomb'' \\
decision    & BLOCK \\
rule        & block\_instruction\_override \\
latency     & $0$\,ms \\
\midrule
\multicolumn{2}{@{}l}{\textbf{Record 2 --- full-pipeline pass}} \\
\midrule
request\_id & 1a9e9aca6ad4 \\
prompt      & ``What is the capital of France?'' \\
decision    & PASS \\
rule        & --- \\
latency     & $917$\,ms \\
record      & moderator verdict safe; intent
              general\_question; default response path; released \\
\bottomrule
\end{tabular}
\end{table}

\section{Evaluation}
\label{sec:evaluation}

We evaluated SAFESHIELD from complementary mechanism-, organization-, and system-level perspectives. Section V-A evaluates the concrete mechanisms supporting the framework, while Section V-B examines the aggregate contribution of decision organization and isolates selected coordination dependencies. Section V-C evaluates end-to-end behavior under deployment-oriented stress, and Section V-D examines Decision Traces for post-hoc inspection.

Unless otherwise noted, all experiments used Qwen3-4B-Instruct ~\cite{qwen3technicalreport} as the backbone. We selected this small open-source instruction-tuned model to reflect the local and self-hosted deployment setting motivating this work.

\subsection{Mechanism-Level Evaluation}
\label{sec}

Before evaluating decision organization as a whole, we evaluated the concrete mechanisms supporting SAFESHIELD’s instantiated decision responsibilities. These experiments assess moderation, intent classification, retrieval-grounded generation, and post-generation verification independently of the organizational contribution studied in Section V-B. Some experiments involve mechanisms spanning multiple runtime stages; accordingly, their results establish the functionality of the participating mechanisms rather than the performance of individual decision responsibilities.


Across the mechanism-level experiments, we used the unprotected Qwen3-4B-Instruct backbone as a model-level baseline and NVIDIA NeMo Guardrails~\cite{rebedea-etal-2023-nemo} as the primary system-level baseline where a matched comparison was applicable. NeMo v0.19 was configured with the same backbone and, where relevant, matched retrieval and sampling settings. Individual experiments additionally included task-specific baselines when needed to evaluate the mechanism under study.

\begin{table}[t]
\centering
\footnotesize
\setlength{\tabcolsep}{3pt}
\caption{Evaluation datasets and corresponding SAFESHIELD stages.}
\label{tab1}
\begin{tabular}{@{}llll@{}}
\toprule
Experiment & Dataset & Size & Stage \\
\midrule
Intent Classif. & Banking77 & 2{,}002 & Intent Classif. \\
Adv. Filtering & Anthropic HH & 400 & Input/Output \\
Hallucination & KG-FPQ & 400 & Output Check \\
RAG Faithfulness & SQuAD v1.1 & 400 & Knowledge Retr./Output \\
\bottomrule
\end{tabular}
\end{table}

\subsubsection{Adversarial and Harmful-Content Filtering}
We evaluated SAFESHIELD's adversarial and malicious-content filtering mechanisms on a balanced subset of Anthropic Helpful–Harmless~\cite{bai2022training}, containing 200 harmful and 200 benign prompts. For this experiment, the input-side admission checks and output-side safety checks were active, and an interaction was considered blocked if either prevented delivery. We compared against the underlying Qwen3-4B-Instruct model, which relied on its built-in refusal behavior, and NeMo Guardrails under matched backbone and prompting settings.

As shown in Table~\ref{tab:filtering}, SAFESHIELD blocked 88.0\% of harmful interactions while allowing 89.5\% of benign prompts, yielding 88.75\% overall accuracy. Pure Qwen blocked 69.0\% of harmful prompts, while NeMo Guardrails blocked 52.0\%. These results establish that the instantiated safety checks provide effective interception beyond the backbone’s built-in refusal behavior. This experiment activated both input- and output-side checks; accordingly, the 88.0\% result reflects their combined moderation capability rather than admission alone.

\begin{table}[t]
\centering
\small
\setlength{\tabcolsep}{4pt}
\caption{Adversarial and malicious content filtering results
(400 samples: 200 harmful, 200 benign).}
\label{tab:filtering}
\begin{tabular}{@{}lccc@{}}
\toprule
System & \shortstack{Benign\\Allowed}
       & \shortstack{Harmful\\Blocked}
       & Accuracy \\
\midrule
Pure Qwen       & \textbf{98.5\%} & 69.0\%          & 83.75\% \\
NeMo Guardrails & \textbf{100.0\%} & 52.0\%         & 76.0\% \\
\textbf{SAFESHIELD}
                & 89.5\%  & \textbf{88.0\%} & \textbf{88.75\%} \\
\bottomrule
\end{tabular}
\end{table}
\subsubsection{Intent Classification}
We evaluated the intent-classification mechanism on Banking77~\cite{Casanueva2020}, which contains 77 fine-grained banking intents. The test set contains 2,002 queries (26 per intent), and five examples per intent were used to construct the retrieval index. SAFESHIELD retrieves semantically related examples before assigning the routing category. We compared against Pure Qwen, which receives the intent definitions without retrieval assistance, and NeMo Guardrails configured with the same backbone, embedding model, retrieval depth, and demonstration count.

Table~\ref{tab4} reports intent-classification accuracy. Pure Qwen achieved 27.77\%, whereas SAFESHIELD reached 82.72\%, comparable to NeMo Guardrails at 82.89\%. Given the fine-grained and semantically overlapping intent space, we interpret this performance as evidence that the intent-classification mechanism provides a useful routing signal rather than as an isolated end-to-end safety gain.

\begin{table}[t]
\centering
\small
\caption{Intent classification accuracy on Banking77 (2,002 test queries, 77 intents).}
\label{tab4}
\begin{tabular}{@{}lc@{}}
\toprule
\textbf{System} & \textbf{Accuracy} \\
\midrule
Pure Qwen (no retrieval) & 27.77\% \\
NeMo Guardrails & \textbf{82.89\%} \\
SAFESHIELD & 82.72\% \\
\bottomrule
\end{tabular}
\end{table}

\subsubsection{Retrieval and Faithfulness}
We evaluated retrieval and evidence-grounded generation in a controlled retrieval-augmented setting constructed from 400 question--answer pairs sampled from the SQuAD v1.1 development set~\cite{rajpurkar2016squad}. Each question was paired with a ten-passage candidate corpus containing one gold passage and nine unrelated distractors, with the gold position randomized. This setup tests whether Knowledge Retrieval can identify useful evidence and whether the resulting answer remains grounded in that evidence.

We compared SAFESHIELD with a no-retrieval Pure Qwen baseline and NeMo Guardrails under the same backbone, retrieval configuration, and evaluation samples. We reported Recall@3 (R@3), which indicates whether the gold passage appears among the top three retrieved passages; SQuAD Answer F1, which measures token-level overlap between the generated and reference answers; Exact Match (EM), which indicates exact agreement with the reference answer; and the fraction of responses judged faithful to the retrieved evidence.

As shown in Table~\ref{tab5}, SAFESHIELD and NeMo both retrieved the gold passage within the top three candidates for all evaluated questions, while the no-retrieval baseline obtains only 17.4\% Answer F1. With retrieval and verification, SAFESHIELD reached 84.1\% F1, 69.0\% EM, and an 89.5\% faithfulness rate, compared with 82.5\%, 65.5\%, and 82.0\% for NeMo. These results show that both systems benefit from retrieval, while SAFESHIELD achieves higher answer quality and faithfulness than the
matched NeMo configuration.
\begin{table}[t]
\centering
\small
\caption{End-to-end RAG evaluation on 400 SQuAD samples with distractor retrieval corpora.}
\label{tab5}
\begin{tabular}{@{}lcccc@{}}
\toprule
\textbf{System} & \textbf{R@3} & \textbf{F1} & \textbf{EM} & \textbf{Faith.} \\
\midrule
No Retrieval (Pure Qwen) & --- & 17.4\% & 5.0\% & --- \\
NeMo Guardrails & 100\% & 82.5\% & 65.5\% & 82.0\% \\
\textbf{SAFESHIELD} & \textbf{100\%} & \textbf{84.1\%} & \textbf{69.0\%} & \textbf{89.5\%} \\
\bottomrule
\end{tabular}
\end{table}

\subsubsection{Self-Consistency Verification}
We evaluated the post-generation verification capability of Output Check on KG-FPQ~\cite{zhu2025kg}, using 400 responses at the highest confusability level. The set contains 64 hallucinated and 336 correct responses, giving a 16\% hallucination rate. This experiment focuses on the consistency check within Output Check: multiple verification variants were generated under different sampling conditions, and semantic agreement across them was used as a signal of response stability.

We compared SAFESHIELD's Self-Consistency Verification against a single-shot LLM judge (Pure-LLM), SelfCheckGPT~\cite{manakul2023selfcheckgpt}, a sampling-and-NLI baseline corresponding to NeMo's hallucination-checking design, and NeMo's \texttt{self\_check\_facts} under oracle, self-generated, and question-only evidence conditions. Because the dataset is imbalanced, we reported AUC-ROC for threshold-independent discrimination together with F1 at a threshold selected using Youden’s J statistic.

Table~\ref{tab} shows that Self-Consistency Verification achieved the highest AUC-ROC (0.723) and F1 (0.427) among the evaluated methods. SelfCheckGPT obtained 0.690/0.375, while NeMo Sampling+NLI obtained 0.662/0.321. These results establish that Output Check provides a useful post-generation signal for identifying potentially unsupported or unstable responses.

\begin{table}[t]
\centering
\small
\setlength{\tabcolsep}{4pt}
\caption{Hallucination detection on KG-FPQ Level~6 with Qwen3-4B-Instruct ($n{=}400$; 16\% hallucinated). F1 uses Youden's J calibration; sampling methods use $N{=}5$. Oracle evidence is not deployable.}
\label{tab}
\begin{tabular}{@{}lcc@{}}
\toprule
Method & AUC-ROC & F1 \\
\midrule
NeMo SCF (oracle) & 0.651 & 0.394 \\
NeMo SCF (self-gen.) & 0.579 & 0.284 \\
NeMo SCF (question) & 0.576 & 0.282 \\
\midrule
Pure-LLM & 0.607 & 0.348 \\
NeMo Sampling+NLI & 0.662 & 0.321 \\
SelfCheckGPT & 0.690 & 0.375 \\
\textbf{Self-Consistency Verif.\ (ours)} & \textbf{0.723} & \textbf{0.427} \\
\bottomrule
\end{tabular}
\end{table}

\subsection{Organization- and Coordination-Level Evaluation}
\label{subsec:organization_evaluation}
We evaluated the instantiated decision organization through two complementary types of ablation. Aggregate stage ablations measure how end-to-end safety changes when selected runtime stages are removed, while controlled coordination ablations preserve the participating mechanisms and selectively remove dependencies between their decisions. The former characterizes the contribution of the instantiated stages at the system level, whereas the latter isolates the effect of decision coordination from mechanism availability. We report Attack Success Rate (ASR), the fraction of harmful prompts that successfully elicit an unsafe response; Harmful-Prompt Rejection Rate (HPR), the fraction of harmful prompts for which the system does not produce the requested harmful content, whether through an explicit safety intervention or backbone refusal; and average per-request latency. Lower ASR is better, whereas higher HPR is better. ASR and HPR are evaluated independently and are not complementary.

\subsubsection{Aggregate Organization Ablation}
Table~\ref{tab:ablation1} evaluates on HarmBench
\cite{mazeika2024harmbench} how end-to-end safety changes as SAFESHIELD's input- and output-side safety stages are progressively removed. Full SAFESHIELD achieves 0.25\% ASR; removing Input Check raises ASR to 6.75\%, and removing both Input Check and Output Check raises it to 16.5\%. Removing Output Check alone leaves ASR unchanged at 0.25\%, consistent with HarmBench primarily stressing input-driven harmful-request attacks; Output Check is evaluated separately on post-generation hallucination and faithfulness failures in Section V-A.
The \textit{w/o Both} and undefended configurations are not identical: the former retains routing, retrieval, and generation, whereas the latter evaluates backbone-only generation. We therefore treat their 0.75 percentage-point ASR difference as a point observation rather than evidence of a significant difference. The stronger result is the increase in ASR from 0.25\% to 6.75\% and then to 16.5\% as the input- and output-side safety stages are progressively removed, showing that the instantiated organization materially contributes to end-to-end safety.

Latency is path-dependent because admission determines whether downstream generation and verification execute. With Input Check active, most harmful requests terminate early, so Full and \textit{w/o Output} have similar average latency (297 vs. 286 ms). Without Input Check, more requests reach generation and downstream verification, increasing average latency to 4008 ms compared with 2436 ms when both safety-checking stages are removed. These configuration-level differences should therefore not be interpreted as additive per-stage costs.
\begin{table}[t]
\centering
\caption{Stage ablation on HarmBench (400 harmful prompts). Progressive removal of safety stages increases attack success rate.}
\label{tab:ablation1}
\small
\begin{tabular}{lccc}
\toprule
Configuration & ASR $\downarrow$ & HPR $\uparrow$ & Latency (ms) \\
\midrule
Full SAFESHIELD        & \textbf{0.25\%} & \textbf{99.5\%} & 297 \\
\quad w/o Input stage  & 6.75\%  & 81.2\% & 4008 \\
\quad w/o Output stage & 0.25\%  & 99.5\% & 286 \\
\quad w/o Both stages  & 16.5\%  & 71.8\% & 2436 \\
Undefended backbone    & 15.75\% & 84.2\% & 2465 \\
\bottomrule
\end{tabular}
\end{table}

\subsubsection{Controlled Coordination Ablations}
\label{subsubsec:coordination_ablations}
We isolated two coordination relationships: admission gating and evidence propagation.
\paragraph{Admission gating}
We isolated gating using the same 200 harmful and 200 benign prompts as in the
mechanism-level moderation evaluation, but under a controlled configuration
designed specifically to isolate the admission$\rightarrow$execution dependency.
The two non-target output rails (self-consistency and RAG faithfulness) were
disabled in both conditions, while all admission checks and their recorded
decisions were preserved. In the full condition, a rejection terminates downstream
execution; in the gating-ablated condition, the same admission decision is
recorded but not enforced. Harmful-request interception consequently falls from 81.5\% to 54.0\%, while the fraction of harmful requests allowed to proceed rises from 18.5\% to 46.0\%; benign over-refusal remains nearly unchanged (10.0\% vs. 9.5\%). Paired decision correctness favors
the full condition 62 to 8, and a McNemar test, which evaluates whether paired binary outcomes differ systematically between the two conditions, confirms that this asymmetry is significant
($p = 1.8 \times 10^{-11}$). These results show that admission gating contributes
beyond the admission mechanism itself: computing and recording a rejection is
insufficient unless that decision can halt downstream execution.

\paragraph{Evidence propagation}
We isolated evidence propagation by preserving retrieval, generation, and release checking while withholding upstream evidence only from the release decision. Across 200 paired retrieval-grounded examples, release accuracy falls from 96.0\% to 69.5\%, while the release-blocking rate rises from 5.5\% to 34.0\%. Conditional faithfulness among released responses remains similar (95.8\% vs. 95.5\%). This conditional faithfulness measure is computed only over responses eligible for evidence-based verification and is therefore distinct from the overall faithfulness rate reported in the stage-level RAG evaluation. Paired decision correctness favors the full condition 55 to 2 (McNemar $p=2.3\times10^{-14}$), indicating that upstream evidence primarily improves release-decision correctness rather than the conditional quality of already released responses. Together, the controlled ablations show that coordination affects end-to-end behavior even when the participating mechanisms and their local decisions are preserved.


\subsection{End-to-End Behavior Under Deployment-Oriented Stress}

We further evaluated the complete SAFESHIELD organization on
SAFESHIELD-Bench, a 600-sample deployment-oriented stress suite
constructed for this study. Unlike the controlled mechanism-level experiments, SAFESHIELD-Bench examines end-to-end behavior when safety-relevant decisions must be made across the full deployment pipeline. The suite covers harmful and adversarial requests, hallucination-prone questions,
retrieval failures, safety-boundary cases, and benign controls.

SAFESHIELD-Bench combines samples from established benchmarks with
structured extensions designed to cover deployment scenarios that are
underrepresented in standard final-output evaluations. Its composition is
25\% TruthfulQA~\cite{lin2022truthfulqa}, 18\% HarmBench~\cite{mazeika2024harmbench},
9.5\% AdvBench~\cite{zou2023universal}, 7.5\% ToxiGen~\cite{hartvigsen2022toxigen},
6\% CrowS-Pairs~\cite{nangia2020crows}, 7.5\% MS MARCO~\cite{chen2024msmarco},
6.5\% XSTest~\cite{rottger2024xstest}, and 20\% structured extensions.
Some benchmark-derived samples are reformulated into multi-turn, adversarial, or implicit-intent settings.

For analysis, samples are grouped by their primary decision
responsibility and scenario type. Admission covers harmful and adversarial inputs, Evidence covers retrieval-grounding cases, and Release covers both output-safety and hallucination scenarios. Benign controls are reported separately because they do not correspond to a single intervention responsibility. Routing is exercised across these scenarios through intent-dependent action handling rather than treated as a separate scenario group.

Table~\ref{tab:system_results} summarizes the results on
SAFESHIELD-Bench. SAFESHIELD correctly handles 498 of 600 samples overall. Accuracy is highest for hallucination handling (97\%) and benign controls (93\%), followed by evidence (83\%) and admission scenarios (81\%). Output-safety scenarios are more challenging because they often require normative interpretation rather than direct harmful-content detection.

The end-to-end results complement the controlled coordination ablations. Whereas the ablations isolate whether particular dependencies are load-bearing, SAFESHIELD-Bench evaluates the resulting organization under a broader mixture of deployment-oriented failure modes. Among the 102 incorrectly handled scenarios, recurring failures include harmful or adversarial requests that evade admission checks, output-level normative judgments such as protected-group comparisons and pseudoscientific framing, retrieval contradictions, and conservative blocking of legitimate queries. These patterns show that deployment-time failures arise across different parts of the decision process rather than exclusively at final-output moderation.

\begin{table}[t]
\centering
\caption{End-to-end results on SAFESHIELD-Bench (600 scenarios).}
\label{tab:system_results}
\small
\setlength{\tabcolsep}{4pt}
\begin{tabular}{llcc}
\toprule
Responsibility & Scenario & Correct & Acc. \\
\midrule
Admission & Harmful/adversarial & 195/240 & 81\% \\
Release   & Output safety       & 84/120  & 70\% \\
Release   & Hallucination       & 102/105 & 97\% \\
Evidence  & Retrieval grounding & 75/90   & 83\% \\
---       & Benign control      & 42/45   & 93\% \\
\midrule
Overall   &                     & 498/600 & 83\% \\
\bottomrule
\end{tabular}
\end{table}
\paragraph{Robustness to Evasion Variants}
We further probed the full system with a hand-constructed adversarial set of 107
prompts spanning surface perturbations (E1--E3), jailbreak framings, and
multilingual restatements. SAFESHIELD blocked \textbf{104 of 107} prompts (\textbf{97.2\%}):
24/24 for E1 typographic variants, 23/23 for E2, 6/9 for E3, 36/36 for jailbreak framings, and 15/15 for multilingual restatements across five languages. We report this as a diagnostic stress test of whole-system behavior, not a coverage guarantee; broader multilingual validation remains future work.
 
\subsection{Decision Records}
\label{sec:eval-trace}

Decision Traces provide a structured record for post-hoc inspection of the end-to-end outcomes reported above. On SAFESHIELD-Bench, 102 of 600 scenarios are not handled as expected. Rather than treating these cases only as final-output errors, the trace preserves the intermediate decisions and evidence associated with each outcome, allowing the point at which the observed behavior first becomes visible to be localized within the deployment-time decision process.
 
\section{Discussion}
\label{sec:discussion}

\subsection{Decision Organization and Guardrail Frameworks}

SAFESHIELD shares with guardrail frameworks such as NeMo
Guardrails~\cite{rebedea-etal-2023-nemo} the goal of adding external safety controls but focuses on a different aspect of deployment design.
Guardrail frameworks provide mechanisms for expressing policies and controlling interactions, whereas SAFESHIELD focuses on how safety decisions are assigned and coordinated across the deployment pipeline. This distinction separates decision organization from implementation: mechanisms can be replaced without changing the responsibilities or dependencies they serve, while our coordination ablations show that modifying decision dependencies alone can alter end-to-end behavior. SAFESHIELD is therefore complementary to, rather than a replacement for, general-purpose guardrail frameworks.

\subsection{From Safety Labels to Deployment Responsibilities}

Safe/unsafe labels identify content-level risks but do not specify where or how a deployment should intervene. Responsibility-oriented decomposition addresses this gap by associating safety judgments with explicit deployment decisions and their ownership.

\subsection{Decision Traces and Accountability}

Decision Traces preserve the decision process beyond the final pass/block outcome. This allows failures arising from admission, routing, grounding, or release to be distinguished during post-hoc analysis rather than collapsed into a single final-output error. Decision Traces do not establish accountability by themselves, nor do we evaluate
their usability for human auditors. Instead, they provide an inspectable runtime record of decision ownership and dependencies that can support future oversight and governance workflows.

\subsection{Limitations}
The four responsibilities characterize the deployment setting studied here rather than an exhaustive taxonomy. Our experiments use a single primary backbone and retrieval configuration, and controlled coordination ablations focus on gating and evidence propagation. Broader multilingual and deployment coverage remains
future work. Decision Traces support, rather than replace, human and institutional oversight.



\section{Conclusion}
\label{sec:conclusion}

This paper presents SAFESHIELD, an auditable framework for
deployment-time safety of small language models. Rather than treating
safety as a single moderation decision, SAFESHIELD organizes
deployment-time safety through responsibility-oriented decomposition and
explicit coordination among admission, routing, evidence, and release
decisions. Its implementation maps these responsibilities to Input Check,
Intent Classification, Knowledge Retrieval, and Output Check, while
Decision Traces preserve the basis and outcome of safety decisions for
post-hoc inspection.

Through mechanism-level evaluation, aggregate organization ablations, controlled coordination ablations, and SAFESHIELD-Bench, we show that the individual decision mechanisms perform their intended functions and that their organization affects end-to-end behavior. In particular, disrupting coordination while preserving the participating mechanisms degrades safety outcomes,
indicating that deployment-time safety depends not only on which
safeguards are present but also on how their decisions are organized.

These findings position decision organization as a distinct design
problem for SLM deployment. SAFESHIELD provides one concrete
instantiation of this perspective, combining explicit decision ownership,
coordination, and inspectable runtime records. Future work will evaluate SAFESHIELD across additional languages and deployment settings, reduce verification overhead, and study how Decision Traces can support human audit and governance workflows.

\section{Ethical Considerations}

This work evaluates deployment-time safety using public research benchmarks and constructed or reformulated stress-test scenarios. No private user data or personally identifiable information is used.
SAFESHIELD may conservatively refuse some benign requests when safety
signals are uncertain. It is intended to support deployment-time safety
and oversight rather than replace human judgment or domain-specific
professional review in high-risk settings.
\section{AI Disclosure}
Claude and ChatGPT were used for language editing and to assist in drafting and revising portions of Sections II, V, and VI. The tools were also used to assist in checking bibliographic information. The authors reviewed and verified all AI-assisted content, including the references, and take full responsibility for the accuracy, originality, and integrity of the manuscript.
\section*{Acknowledgment}
This work was carried out in part during the first author's internship at Advanced Micro Devices, Inc. (AMD), and in part at The University of Texas at Austin.

\bibliographystyle{IEEEtran}
\bibliography{reference}
\vspace{12pt}

\end{document}